\documentclass
[superscriptaddress,secnumarabic,amssymb,amsmath,nobibnotes,aps,prd,showkeys,showpacs,nofootinbib,onecolumn]{revtex4}%
\usepackage{graphicx}
\usepackage{epsf}
\usepackage{bm}
\usepackage{amsmath}
\usepackage{amsfonts}
\usepackage{amssymb}
\usepackage{epstopdf}%
\providecommand{\U}[1]{\protect\rule{.1in}{.1in}}

\newcommand{\be}{\begin{equation}}
\newcommand{\ee}{\end{equation}}

\newcommand{\mincir}{\raise
-3.truept\hbox{\rlap{\hbox{$\sim$}}\raise4.truept\hbox{$<$}\ }}
\newcommand{\magcir}{\raise
-3.truept\hbox{\rlap{\hbox{$\sim$}}\raise4.truept\hbox{$>$}\ }}

\begin{document}
\title{Cosmological Solutions of $f(T)$ Gravity}
\author{Andronikos Paliathanasis}
\email{anpaliat@phys.uoa.gr}
\affiliation{Instituto de Ciencias F\'{\i}sicas y Matem\'{a}ticas, Universidad Austral de
Chile, Valdivia, Chile}
\author{John D. Barrow}
\email{J.D.Barrow@damtp.cam.ac.uk}
\affiliation{DAMTP, Centre for Mathematical Sciences, University of Cambridge, Wilberforce
Rd., Cambridge CB3 0WA, UK}
\author{P.G.L. Leach}
\email{leach.peter@ucy.ac.cy}
\affiliation{Department of Mathematics and Institute of Systems Science, Research and
Postgraduate Support, Durban University of Technology, PO Box 1334, Durban
4000, Republic of South Africa}
\affiliation{School of Mathematics, Statistics and Computer Science, University of
KwaZulu-Natal, Private Bag X54001, Durban 4000, Republic of South Africa}
\affiliation{Department of Mathematics and Statistics, University of Cyprus, Lefkosia 1678, Cyprus}

\begin{abstract}
In the cosmological scenario in $f\left(  T\right)  $ gravity, we find
analytical solutions for an isotropic and homogeneous universe containing a
dust fluid and radiation and for an empty anisotropic Bianchi I universe. The
method that we apply is that of movable singularities of differential
equations. For the isotropic universe, the solutions are expressed in terms of
a Laurent expansion, while for the anisotropic universe we find a family of
exact Kasner-like solutions in vacuum. Finally, we discuss when a nonlinear
$f\left(  T\right)  $-gravity theory provides solutions for the teleparallel
equivalence of general relativity and derive conditions for exact solutions of
general relativity to solve the field equations of an $f(T)$ theory.

\end{abstract}
\maketitle
\date{\today}

\section{Introduction}

One of the most important unsolved problems of modern astronomy and particle
physics is the identity of the 'dark energy' that is evidently responsible for
the observed acceleration of the universal expansion. The dynamics can be
quite accurately described by the inclusion of a simple cosmological constant
term to Einstein's equations but its required magnitude is mysterious and
unmotivated by fundamental physics. Some new fundamental theory might
eventually be able to provide a natural explanation (see, for an example,
\cite{bshaw}) or there may be more complicated explanations in which the dark
energy is not a constant stress, but some time-dependent scalar (or effective
scalar) field. Effective scalar fields are available in the many deviant
theories of gravity that have been proposed as generalisations of Einstein's
general theory of relativity. In the past it had been expected that deviations
from Einstein's theory would only arise in situations of high, or formally
infinite, spatial curvature -- so called curvature 'singularities' -- where
the entire theory breaks down. However, the unusual challenge posed by the
acceleration of the universe is that it may require modifications to
Einstein's theory in the late universe when spatial curvature is very low. So
called 'modified theories of gravity' provide one of these scenarios to
explain the acceleration of the universe. In contrast to the explicit dark
energy models, such as quintessence, phantom fields, Chaplygin gas or many
others (see \cite{Ratra88,Lambda4,Brookfield2005td,aasen} and references
therein), in which an energy-momentum tensor which violates the strong energy
condition is added to the field equations of General Relativity (GR), in
modified gravity theories the dark energy often has a geometric origin and is
related to new dynamical terms which follow from the modification of the
Einstein-Hilbert action.

A particular modified theory of gravity which has attracted the interests of
cosmologists is so-called $f\left(  T\right)  ~$teleparallel
gravity\footnote{For a recent review on $f\left(  T\right)  $-gravity see
\cite{ftRev}.} \cite{Ferraro,Ferraro06,Lin2010}. Inspired by the formulation
of $f\left(  R\right)  $-gravity, in which the lagrangian of the gravitational
field equations is a function, $f$, of the Ricci scalar $R$ of the underlying
geometry, $f\left(  T\right)  $ gravity is a similar generalization. Now,
instead of using the torsionless Levi-Civita connection of GR, the
curvatureless Weitzenb{\"{o}}ck connection is used in which the corresponding
dynamical fields are the four linearly independent vierbeins, and $T$ is
related to the antisymmetric connection which follows from the non-holonomic
basis \cite{Hayashi79,Tsamp,ft1}.

A linear $f\left(  T\right)  $-theory leads to the teleparallel equivalent of
GR (TEGR) \cite{ein28}. However, $f\left(  T\right)  $ gravity does not
coincide with $f\left(  R\right)  $ gravity. One of the main differences is
that for a nonlinear $f\left(  R\right)  $ function, gravity is a fourth-order
theory, whereas $f\left(  T\right)  $-gravity is always a second-order theory.
This follows because $T$ includes only first derivatives of the vierbeins.
Moreover, while $f\left(  T\right)  $-gravity is a second-order theory and in
the limit of a linear function, $f=R,$~GR is recovered, in general $f\left(
T\right)  $-gravity provides different structural properties from those of GR.
However, from the analysis of the cosmological data and the solar system tests
of GR we know that deviations from GR must be small, and so $f(T)$ must be
close to a linear form \cite{Wu2010,Bengochea001,Iorio2,Bas}.

Even though $f\left(  T\right)  $-gravity is a second-order theory, very few
exact analytical solutions of the field equations are known. Some power-law
solutions in a Friedmann-Lema\^{\i}tre-Robertson-Walker (FLRW) spacetime can
be found in \cite{Ataz,palft}, while some power-law solutions in anisotropic
spacetimes are given in \cite{an1}. Finally, some analytical solutions in the
case of static spherically symmetric spacetimes can be found in
\cite{palft2,st1}, and references therein.

In this work we are interested to determine exact solutions of the field
equations in $f\left(  T\right)  $-gravity in the cosmological scenario of an
isotropic and homogeneous universe and for the Bianchi I spacetime.
Specifically, the method that we use is that of the singularity analysis of
differential equations. Singularity analysis is complementary to symmetry
analysis (for a discussion between the two methods see \cite{palLeach2}). The
application of Noether point symmetries for $f\left(  T\right)  $-gravity can
be found in \cite{Ataz,palft,palft2}. \ Recently, singularity analysis was
applied in the cosmological scenario of $R+\alpha R^{n}$ gravity
\cite{palLeach} and it was proved that, if $n$ is a rational number and $n>1$,
then the gravitational field equations pass the singularity test and the
analytical solution of the field equations can be written as a Laurent
expansion around the movable singularity of the field equations. The
application of singularity analysis in gravitational studies is not new and
has provided interesting results \cite{CotsakisLeach,Con1,Demaret,Cotsakis}.

The plan of the paper is as follows. In Section \ref{field} we define our
model with $f\left(  T\right)  $-gravity in a spatially flat FLRW spacetime
and the resulting gravitational field equations are presented. The singularity
analysis of the field equations for some functions proposed in the literature
for $f\left(  T\right)  $ is performed in Sections \ref{sing} and
\ref{bianchi1}. Specifically, we consider the power-law model $f\left(
T\right)  =T+\alpha\left(  -T\right)  ^{n},$ which has been proposed in
\cite{Ferraro} as was the same model with the cosmological constant term:
$f\left(  T\right)  =T+\alpha\left(  -T\right)  ^{n}-\Lambda.$ For these two
models we find that the solution of the field equations for the FLRW universe
can be written analytically in a Laurent expansion. However, the singularity
analysis fails in the Bianchi I spacetime, but in the latter model we find
that there exists an exact vacuum solution of the field equation which leads
to a Kasner-like universe. In Section \ref{grsol}, we construct conditions
which allow solutions of GR to be recovered in $f\left(  T\right)  $-gravity.
Finally, in Section \ref{conc}, we discuss our results and draw our conclusions.

\section{ \ $f\left(  T\right)  $-gravity}

\label{field}

We briefly discuss the basic assumptions of $f(T)$ teleparallel gravity. The
vierbein fields, ${\mathbf{e}_{i}(x^{\mu})~}$, as non-holonomic frames in
spacetime, are the dynamical variables of teleparallel gravity and
consequently of the $f\left(  T\right)  $ gravity. The vierbein fields form an
orthonormal basis for the tangent space at each point $x^{\mu}$ of the
manifold, that is, $g(e_{i},e_{j})=\mathbf{e}_{i}\cdot\mathbf{e}_{i}=\eta
_{ij}$, where $\eta_{ij}~$is the line element of four-dimensional Minkowski
spacetime. In a coordinate basis~the vierbeins can be written as $e_{i}%
=h_{i}^{\mu}\left(  x\right)  \partial_{i},$ for which the the metric tensor
is defined as follows%
\begin{equation}
g_{\mu\nu}(x)=\eta_{ij}h_{\mu}^{i}(x)h_{\nu}^{j}(x). \label{ft.01}%
\end{equation}

The curvatureless Weitzenb\"{o}ck connection, which is considered in
teleparallel gravity, has the non-null torsion tensor \cite{Haya,Maluf},
\begin{equation}
T_{\mu\nu}^{\beta}=\hat{\Gamma}_{\nu\mu}^{\beta}-\hat{\Gamma}_{\mu\nu}^{\beta
}=h_{i}^{\beta}(\partial_{\mu}h_{\nu}^{i}-\partial_{\nu}h_{\mu}^{i}),
\label{ft.02}%
\end{equation}
while the lagrangian density of the teleparallel gravity, from which the
gravitational field equations are derived, is the scalar,%

\begin{equation}
T={S_{\beta}}^{\mu\nu}{T^{\beta}}_{\mu\nu}, \label{ft.02a}%
\end{equation}
$~$

where
\begin{equation}
{S_{\beta}}^{\mu\nu}=\frac{1}{2}({K^{\mu\nu}}_{\beta}+\delta_{\beta}^{\mu
}{T^{\theta\nu}}_{\theta}-\delta_{\beta}^{\nu}{T^{\theta\mu}}_{\theta})
\label{ft.03}%
\end{equation}
and ${K^{\mu\nu}}_{\beta}$ is the contorsion tensor that is defined by
\begin{equation}
{K^{\mu\nu}}_{\beta}=-\frac{1}{2}({T^{\mu\nu}}_{\beta}-{T^{\nu\mu}}_{\beta
}-{T_{\beta}}^{\mu\nu}), \label{ft.04}%
\end{equation}
It equals the difference between the Levi-Civita connections in the holonomic
and the non-holonomic frame.

The action for $f\left(  T\right)  $ gravity is%
\begin{equation}
S_{f\left(  T\right)  }=\frac{1}{16\pi G}\int d^{4}xe\left(  f(T)\right)
+S_{m}, \label{ft.05}%
\end{equation}
in which $e=\det(e_{\mu}^{i})=\sqrt{-g}$. Variation with respect to the
vierbein gives the gravitational field equations:
\begin{align}
&  e^{-1}\partial_{\mu}(ee_{i}^{\rho}S_{\rho}{}^{\mu\nu})f_{T}-e_{i}^{\lambda
}T^{\rho}{}_{\mu\lambda}S_{\rho}{}^{\nu\mu}f_{T}\nonumber\\
&  \ \ \,+e_{i}^{\rho}S_{\rho}{}^{\mu\nu}\partial_{\mu}({T})f_{TT}+\frac{1}%
{4}e_{i}^{\nu}f({T})=4\pi Ge_{i}^{\rho}\mathcal{T}_{\rho}{}^{\nu},
\label{ft.06}%
\end{align}
where $f_{T}$ and $f_{TT}$ denote the first and second derivatives of the
function $f(T)$ with respect to $T,$ and the tensor $\mathcal{T}_{\rho}{}%
^{\nu}$ denotes the energy-momentum tensor of the matter source~$S_{m}$.
Furthermore, from (\ref{ft.06}) we recover GR when $f_{TT}=0$.

\subsection{Modified Friedmann's equations}

In order to recover the cosmological scenario of a spatially-flat FLRW
spacetime, we consider the diagonal frame for the vierbein:
\begin{equation}
h_{\mu}^{i}(t)=diag(1,a(t),a(t),a(t)). \label{ft.07}%
\end{equation}
In the holonomic frame the spacetime has the line-element%
\[
ds^{2}=-dt^{2}+a^{2}(t)(dx^{2}+dy^{2}+dz^{2}),
\]
where $a(t)$ is the cosmological scale factor. \ For this frame we calculate
the lagrangian density,
\begin{equation}
T=-6\left(  \frac{\dot{a}}{a}\right)  ^{2}=-6H^{2}, \label{ft.08}%
\end{equation}
where $H=\dot{a}/a$ is the Hubble parameter, while the gravitational field
equations (\ref{ft.06}) become
\begin{equation}
12H^{2}f_{T}(T)+f(T)=16\pi G\rho, \label{ft.09}%
\end{equation}
and
\begin{equation}
48H^{2}\dot{H}f_{TT}(T)-4(\dot{H}+3H^{2})f_{T}(T)-f(T)=16\pi Gp, \label{ft.10}%
\end{equation}
in which $\rho~\,$and $p~$ denote the energy density and pressure,
respectively, of the energy-momentum tensor, $\mathcal{T}_{\rho}{}^{\nu}$,
from which we have the conservation equation%

\begin{equation}
\dot{\rho}+3H(\rho+p)=0. \label{ft.10a}%
\end{equation}

\ However, equations (\ref{ft.09}) and (\ref{ft.10}) can be rewritten as
\begin{equation}
H^{2}=\frac{8\pi G}{3}(\rho+\rho_{T}) \label{ft.11}%
\end{equation}
and
\begin{equation}
2\dot{H}+3H^{2}=-8\pi G(p+p_{T}), \label{ft.12}%
\end{equation}
where $\rho_{T}$ and $p_{T}$ are the effective energy density and pressure of
the geometric fluid which follow from the modification of the gravitational
action integral. Specifically, $\rho_{T}$ and $p_{T}$ depend upon $T$ and
$f_{T}$ and are
\begin{equation}
\rho_{T}=\frac{1}{16\pi G}[2Tf_{T}(T)-f(T)-T], \label{ft.13}%
\end{equation}
and%
\begin{equation}
p_{T}=\frac{1}{16\pi G}\left[  4\dot{H}\left(  2Tf_{TT}(T)+f_{T}(T)-1\right)
\right]  -\rho_{T}. \label{ft.14}%
\end{equation}
An effective equation of state parameter for the geometric fluid can be
defined as usual by
\begin{equation}
w_{T}\equiv\frac{p_{T}}{\rho_{T}}=-1+\frac{4\dot{H}[2Tf_{TT}(T)+f_{T}%
(T)-1]}{2Tf_{T}(T)-f(T)-T}\;. \label{ft.15}%
\end{equation}
From this, if we consider that $f\left(  T\right)  =T+F\left(  T\right)  $,
then (\ref{ft.15}) takes the simpler form
\begin{equation}
w_{T}=-\frac{F-TF_{T}+2T^{2}F_{TT}}{\left(  1+F_{T}+2TF_{TT}\right)  \left(
F-2TF_{T}\right)  }. \label{ft.16}%
\end{equation}

In \cite{Bengochea001}, the model \ $f_{I}\equiv f\left(  T\right)
=T+\alpha\left(  -T\right)  ^{n}$ has been proposed as an alternative to the
dark-energy models and fits some of the cosmological data quite well.
Furthermore, the parameters of that model have been derived from cosmography
in \cite{capcosm}, while in \cite{Iorio2} it has been constrained within the
solar system and it has been found that the perturbation to GR solution is
given in terms of powers $r^{2-2n}$ of distance $r$ from a central point mass.
Furthermore, in \cite{Iorio2}, they performed the analysis by including the
cosmological constant term, \textit{ie},$~f_{II}\equiv$ $f\left(  T\right)
=T+\alpha\left(  -T\right)  ^{n}-\Lambda$. These two models,~$f_{I}$~and
$f_{II}$, are the models we study here. In what follows we will consider the
two models $f_{I}\left(  T\right)  $ and $f_{II}\left(  T\right)  $, with
$n\neq0,1$ (as we are in the teleparallel equivalence of GR) and $n\neq
\frac{1}{2}$ (so we are close to GR in the limit). In order to check the
latter condition, consider $f\left(  T\right)  =F\left(  T\right)  +\beta
\sqrt{-T}$, in (\ref{ft.05}), where $F(T)$ is an arbitrary function. Then,
\begin{equation}
S=\frac{1}{16\pi G}\int d^{4}xeF(T)+\frac{\beta}{16\pi G}\int d^{4}xe\sqrt
{-T}, \label{ft.16a}%
\end{equation}
where, using (\ref{ft.08}),\ the second term becomes a total derivative,
\textit{ie}, $e\sqrt{-T}=a^{2}\dot{a}=\frac{1}{3}\frac{d}{dt}\left(
a^{3}\right)  $, which does not affect the field equations. Furthermore, if we
consider that $F\left(  T\right)  =-\beta\Lambda$, then the action
(\ref{ft.05}) with the use of (\ref{ft.08}) becomes%
\begin{equation}
S=\frac{\beta}{16\pi G}\int d^{4}x\left(  a^{2}\dot{a}-a^{3}\Lambda\right)
=-\frac{\beta}{16\pi G}\int d^{4}x\left(  a^{3}\Lambda\right)  .
\label{ft.16b}%
\end{equation}
Hence. the field equations (\ref{ft.08})-(\ref{ft.10}) cannot be recovered.
Consider the diagonal frame%
\begin{equation}
h_{\mu}^{i}(t)=diag(a^{-3}\left(  \tau\right)  ,a(\tau),a(\tau),a(\tau)),
\label{ft.16c}%
\end{equation}
where the line element is that of FLRW spacetime with a lapse function
$N\left(  \tau\right)  =a^{-3}\left(  \tau\right)  $, \textit{ie},
$dt=N\left(  \tau\right)  d\tau$.$~$Again, $\sqrt{-T}$ is a linear function of
$\dot{a},~$and the gravitational lagrangian is a total derivative, which is
something that has not been observed recently in \cite{sar11}.

However, in the case of vacuum, equation (\ref{ft.09}) can be written as
\begin{equation}
f-2Tf_{T}=0, \label{ft.23}%
\end{equation}
which indeed admits as a solution the case $f\left(  T\right)  =\sqrt{-T}$,
but also has a special solution the $f\left(  T\right)  |_{T->0}=0$, which
means that $a\left(  t\right)  =const$, and we have the solution of GR in
empty spacetime.

For the fluid components of the field equations, we take a dust fluid, with
$p_{m}=0$, and a radiation fluid, $p_{r}=\frac{1}{3}\rho_{r}$. We assume that
the two fluids are not interacting and are minimally coupled to gravity, hence
(\ref{ft.10a}) for each fluid gives $\rho_{m}=\rho_{m0}a^{-3}$ and $\rho
_{r}=\rho_{r0}a^{-4}$. At this point we should mention that eq. (\ref{ft.10}),
which is a second-order equation with respect to the scale factor, still \ has
to be solved and the solution is constrained by the first modified Friedmann's
equation (\ref{ft.09}).

\subsection{Anisotropic Bianchi I spacetime}

The second scenario that we consider in this work is the determination of an
analytical solution in a Bianchi I spacetime. To do that we consider the
diagonal frame
\begin{equation}
h_{\mu}^{i}(t)=diag(1,a(t),b(t),c(t)), \label{ft.b01}%
\end{equation}
where the line element is that of Bianchi I spacetime with unknown scale
factors $a(t),b(t)$ and $c(t):$%
\begin{equation}
ds^{2}=-dt^{2}+a^{2}(t)dx^{2}+b\left(  t\right)  dy^{2}+c\left(  t\right)
dz^{2}. \label{ft.b02}%
\end{equation}

The lagrangian density for (\ref{ft.b01}) is
\begin{equation}
T=-\frac{2}{abc}\left(  c\dot{a}\dot{b}+b\dot{a}\dot{c}+a\dot{b}\dot
{c}\right)  \label{ft.b03}%
\end{equation}
from which we can see that (\ref{ft.08}) is recovered in the isotropic
scenario, $a\left(  t\right)  =b\left(  t\right)  =c\left(  t\right)  ,$ which
is the spatially-flat FLRW universe. \ 

With the use of a Lagrange multiplier in (\ref{ft.05}) the lagrangian of the
field equations can be constructed:%
\begin{equation}
L\left(  a,b,c,\dot{a},\dot{b},\dot{c},T\right)  =2f_{,T}\left(  c\dot{a}%
\dot{b}+b\dot{a}\dot{c}+a\dot{b}\dot{c}\right)  +abc\left(  f_{,T}T-f\right)
, \label{ft.b04}%
\end{equation}
where we have assumed that there is no other matter source.

The gravitational field equations are the Euler-Lagrange equations with
respect to the variables $a,~b~$\ and $c,$ equation (\ref{ft.b03}), which
follow from $\frac{\partial L}{\partial T}=0$, and the constraint equation%
\begin{equation}
2f_{,T}\left(  c\dot{a}\dot{b}+b\dot{a}\dot{c}+a\dot{b}\dot{c}\right)
-abc\left(  f_{,T}T-f\right)  =0. \label{ft.b05}%
\end{equation}
This can be derived from the variation of the lapse function $N$, when
$dt=N\left(  \tau\right)  d\tau$, where we have assumed that $N\left(
t\right)  =1$. \ For the spacetime (\ref{ft.b01}), we perform our analysis for
the same models $f_{I}\left(  T\right)  $,~$f_{II}\left(  T\right)  $
introduced explicitly in the last section.

\section{Analytical solutions in FLRW spacetimes}

\label{sing}

In order to determine the analytic solution of the field equations we apply
the method of singularity analysis and we follow the ARS algorithm
\cite{Abl1,Abl2,Abl3}, which is based upon the existence of movable
singularities for the differential equations and is in the spirit of the
approach of Kowalevskaya \cite{Kowa}. We refer the reader to the following
works for the basic properties of the singularity analysis:
\cite{buntis,Feix97a,Andriopoulos06a}.

We perform our analysis for the two different models, $f_{I}$, and $f_{II}$,
that we discussed above for the two cases for the fluid terms: (a) dust and
(b) dust plus radiation.

\subsection{Dust fluid}

The analyses for the two different models with only a dust fluid present are
as follows:

\subsubsection{Model $f_{I}\left(  T\right)  $}

We substitute $a\left(  \tau\right)  =a_{0}\tau^{\sigma}$ in (\ref{ft.10}) and
we search for the dominant terms in order to determine the power $\sigma$.
Note that $\tau=\left(  t-t_{0}\right)  $ and $t_{0}$ is the position of the
singularity. We have two different possibilities, $n<1$, and $n>1$~with
$n\neq\frac{1}{2}$. Note that $n=1$ is the special case of teleparallel GR.

\paragraph{Case $n<1:$}

For values of $n$ smaller than one we find the dominant behaviour,
$\sigma=\frac{2}{3}$ for $a_{0}$ an arbitrary value. That means that the
singularity of the differential equation is that when $a\left(  t_{0}\right)
\rightarrow0$, while in the same time $\dot{a}\left(  t_{0}\right)
\rightarrow\infty$. In order to determine the position of the resonances we
substitute $a\left(  \tau\right)  =a_{0}\tau^{\frac{2}{3}}+m\tau^{\frac{2}%
{3}+s}$ in (\ref{ft.10}), linearize around the $m\simeq0$ and solve the
remaining polynomial which follows from the dominant terms determining $s$.
The polynomial is $s\left(  s+1\right)  =0$, which gives the two solutions
$s_{1}=-1$, and $s_{2}=0$. The value of $s_{1}$ is essential for the existence
of the singularity and gives a check that our analysis is correct. The second
resonance gives us the position of the second constant of integration which is
at the dominant term. Recall that one constant of integration is the position
of the singularity $t_{0}$. Furthermore, as the dominant term is not a
solution of (\ref{ft.10}) because there are remainder terms, the solution is
expressed in a right Painlev\'{e} series with a step $\frac{1}{3}$, so%
\begin{equation}
a\left(  \tau\right)  =a_{0}\tau^{\frac{2}{3}}+%
{\displaystyle\sum\limits_{N=1}^{+\infty}}
a_{N}\tau^{\frac{2+N}{3}}. \label{ft.17}%
\end{equation}

In the solution, (\ref{ft.17}), the only arbitrary constants are the position
of the singularity, $t_{0}$, and the coefficient $a_{0}$. The coefficients,
$a_{N}$, have to be determined from (\ref{ft.10}) and (\ref{ft.09}).

First consider the case $n=-1$. We substitute the solution, (\ref{ft.17}),
into (\ref{ft.10}) and (\ref{ft.09}) which gives $16\pi G\rho_{m0}=\frac{8}%
{3}a_{0}^{3}$. The non-zero coefficients, $a_{N}$, are the~$a_{M}%
~$with$~M=12\lambda~,~\lambda\in%
\mathbb{N}
$ and $\frac{a_{12}}{a_{0}}=-\frac{9}{320}\alpha$, $~a_{24}=\frac{33}%
{160}\alpha a_{12}$, $a_{36}=\frac{23373}{45760}\alpha a_{24}$ \textit{etc},
occur every twelve terms.

Since there are so many zero coefficients of the $a_{N}$ very close to the
singularity at $a\left(  t_{0}\right)  $, the solution of the field equation
is well approximated by the power-law solution $a\left(  \tau\right)
=a_{0}\tau^{\frac{2}{3}}$, which is that of the dust fluid. That means that
close to the singularity the dominant term in the gravitational field
equations is the linear term $T$, while the dynamical parts contributed by
$T^{n}$ only change the dynamics far from the movable singularity.

\paragraph{Case $n>1:$}

For $n>1,$ the dominant term is $a\left(  \tau\right)  =a_{0}\tau^{\frac{2}%
{3}n}$. We assume that $\frac{2}{3}n\notin%
\mathbb{N}
^{\ast}$ and we calculate that the dominant terms are $\tau^{-2+\frac{2}%
{3}n+s},$ which gives the resonances $s=-1$, $s=0$, so as before the solution
is expressed in a right Painlev\'{e} series. In contrast to the~ $n<1$ case,
$n$ now has to be a rational number in order for the singularity analysis to
work. The step of the right Painlev\'{e} series depends on $n$ and is
determined from the denominator of the dominant term with $\sigma=\frac{2}%
{3}n$.

On the other hand, when $n=\frac{3}{2}\mu,~\mu\in%
\mathbb{N}
^{\ast}$, in order to perform the singularity analysis we substitute
$a\rightarrow b^{-1}\left(  \tau\right)  $, from which we see that the
dominant behaviour is $b\left(  \tau\right)  =b_{0}\tau^{-\mu}$. The
resonances are again at $s=-1$ and$~s=0$ but, as the dominant behaviour is not
a solution of the field equations, the solution is expressed again as a right
Painlev\'{e} series with step one.

Now consider the case $n=2$. The analytical solution is%
\begin{equation}
\frac{a\left(  \tau\right)  }{a_{0}}=\tau^{\frac{4}{3}}+%
{\displaystyle\sum\limits_{N=1}^{+\infty}}
a_{N}\tau^{\frac{4+N}{3}}, \label{ft.18}%
\end{equation}
where the only non-zero coefficients are the $a_{\Sigma}$ with $\Sigma
=6\lambda,~\lambda\in%
\mathbb{N}
$. The constant of integration is $a_{0}$. For the leading coefficients we
have $a_{6}=\left(  288\alpha\right)  ^{-1}~,~a_{12}=17\left(  2880\alpha
\right)  ^{-1}a_{6},~a_{18}=835\left(  205632\alpha\right)  ^{-1}a_{12}$,
\textit{etc} and $16\pi G\rho_{m0}=-\frac{1024}{3}a_{0}^{3}\alpha$, which
means that $\alpha<0$ for $\rho_{m0}>0$. We can see that the solution
(\ref{ft.18}) passes the consistency test. Before we proceed to our analysis
for the second model, $f_{II}\left(  T\right)  $, we note that the dominant
term follows from the $\left(  -T\right)  ^{n}$ term of the action and it is
the power solution of the power-law model $f\left(  T\right)  =\left(
-T\right)  ^{n}~$\cite{palft}; that is, the universe is dominated by the
geometric effective fluid $\rho_{T},~p_{T}$. The fluid has a constant equation
of state parameter $w_{T}=\frac{n-1}{n}$ which is always positive for $n>1$.

On the other hand, for $n=\frac{3}{2}$, which means $\mu=1$, the solution for
the scale factor is
\begin{equation}
\left(  a\left(  \tau\right)  \right)  ^{-1}=b_{0}\tau^{-1}+%
{\displaystyle\sum\limits_{N=1}^{+\infty}}
b_{N}\tau^{-1+N}. \label{ft.19}%
\end{equation}
For the coefficients $b_{N},$ we have the relations: $\frac{b_{1}}{b_{0}%
}=\left(  12\sqrt{6}\alpha\right)  ^{-1}$,~$\frac{b_{2}}{b_{0}}=-\left(
12\sqrt{6}\alpha\right)  ^{-1}\frac{b_{1}}{b_{0}}~$,~$\frac{b_{3}}{b_{0}%
}=\left(  9\left(  12\sqrt{6}\right)  ^{2}\alpha^{2}\right)  ^{-1}\frac{b_{1}%
}{b_{0}}$~,~$\frac{b_{4}}{b_{0}}=\left(  \frac{15}{19}\left(  12\sqrt
{6}\right)  ^{4}\alpha^{4}\right)  ^{-1}\frac{b_{1}}{b_{0}}$ \textit{etc}%
,~while (\ref{ft.09}) gives $16\pi G\rho_{m0}=\frac{12\sqrt{6}}{b_{0}^{3}%
}\alpha>0$. From (\ref{ft.19}), we observe that near the singularity the
effective fluid is that of radiation. We continue our analysis with the model
$f_{II}\left(  T\right)  $ in which the cosmological constant is considered.

\subsubsection{Model $f_{II}\left(  T\right)  $}

The singularity analysis for $f_{II}\left(  T\right)  $ provides the same
results as that of $f_{I}\left(  T\right)  $. This means that the cosmological
constant term does not effect the dominant behaviour near the singularity or
the resonances. The only differences which arise are that the coefficient
terms of the Laurent expansion now also depend upon $\Lambda$. We demonstrate
this by deriving the coefficients for the cases $n=-1,~n=2$ and $n=\frac{3}%
{2}$.

For $n=-1,$ the solution of the field equations for $f_{II}\left(  T\right)  $
is again given by (\ref{ft.17}), where the non-zero coefficients are now
$a_{\bar{M}}~~$with$~\bar{M}=6\lambda~,~\lambda\in%
\mathbb{N}
$. In the analysis above the non-zero coefficients occurred every twelve
steps. The values of the first coefficients are now%

\[
\left(  \frac{a_{6}}{a_{0}}\right)  =\frac{\Lambda}{24}~,~\left(  \frac
{a_{12}}{a_{0}}\right)  =\left(  \frac{\Lambda^{2}-81\alpha}{2880}\right)
~\ \text{and}\left(  \frac{a_{18}}{a_{0}}\right)  =\frac{\Lambda\left(
\Lambda^{2}-1994\alpha\right)  }{362880}.
\]
$~$

Thus, we can see, for $\Lambda=0,$ that the coefficients have the values of
the model $f_{I}\left(  T\right)  _{|n\rightarrow-1}$. Note that we
have~$16\pi G\rho_{m0}=\frac{8}{3}a_{0}^{3}$.

In the case when $n=2$, the solution of field equation is the right
Painlev\'{e} series, (\ref{ft.18}). The non-zero coefficients are~$a_{\Sigma}%
$, with $\Sigma=6\lambda,~\lambda\in%
\mathbb{N}
,$ where the first coefficients are~%

\[
\left(  \frac{a_{6}}{a_{0}}\right)  =\left(  288\alpha\right)  ^{-1},~\left(
\frac{a_{12}}{a_{0}}\right)  =10\alpha\left(  17-162\alpha\Lambda\right)
\left(  \frac{\alpha_{6}}{a_{0}}\right)  ^{2}\text{ and }\left(  \frac{a_{18}%
}{a_{0}}\right)  =84\left(  167-1944\alpha\Lambda\right)  \left(  \frac
{\alpha_{6}}{a_{0}}\right)  ^{3},etc.
\]

Hence, we can see that the cosmological constant affects the dynamics from the
twelfth term of the Laurent expansion and for $\Lambda=0$ we have the same
coefficients as before. Furthermore,\ the first Friedmann equation gives
$16\pi G\rho_{m0}=-\frac{1024}{3}a_{0}^{3}\alpha$.

Finally, for the case of $n=\frac{3}{2}$, the solution of the field equations
is (\ref{ft.19}), where from (\ref{ft.09}) we have $16\pi G\rho_{m0}%
=\frac{12\sqrt{6}}{b_{0}^{3}}\alpha\,$\ and from (\ref{ft.10}) that
\[
\left(  \frac{b_{1}}{b_{0}}\right)  =\left(  12\sqrt{6}\alpha\right)
^{-1},~\left(  \frac{b_{2}}{b_{0}}\right)  =-\left(  12\sqrt{6}\alpha\right)
^{-1}\left(  \frac{b_{1}}{b_{0}}\right)  \text{ and }\frac{b_{3}}{b_{0}}%
=\frac{1-54\alpha^{2}\Lambda}{7776\sqrt{6}\alpha^{3}}.
\]
\textit{ }

From these coefficients we can see that, when $\Lambda=0$, the solution
reduces to that of the model $f_{I}\left(  T\right)  _{|n\rightarrow\frac
{3}{2}}$.

\subsection{Dust and radiation fluids}

In a more general scenario we assume that the matter source of the field
equations includes a part from the cold dark matter (dust) a radiation
component. We use the model $f_{II}\left(  T\right)  $, because the
cosmological constant does not effect the dominant term or the resonances.
Again, we consider two possible cases, $n<1$ and $n>1$.

\paragraph{Case $n<1:$}

We follow the same steps as before and we find that the dominant term of
equation (\ref{ft.10}) is $a\left(  \tau\right)  =a_{0}\tau^{\frac{1}{2}}$.
Now $a_{0}~$is not arbitrary as above, but $\bar{\rho}_{r0}=\frac{1}{2}\left(
a_{0}\right)  ^{4}$,~where $\bar{\rho}_{r0}=$ $\frac{16\pi G}{3}\rho_{r0}$.
This means that the radiation fluid dominates in the early universe as
expected. For the resonances, we find that they are $s_{1}=-1$,~$s_{2}%
=\frac{1}{2}$ and now the position of the second constant of integration in
the Laurent expansion is in the coefficient $a_{1}$. The Laurent expansion is
a right Painlev\'{e} series and is%
\begin{equation}
a\left(  \tau\right)  =a_{0}\tau^{\frac{1}{2}}+a_{1}\tau+%
{\displaystyle\sum\limits_{N=2}^{+\infty}}
a_{N}\tau^{\frac{1+N}{2}}. \label{ft.20}%
\end{equation}

In this case it is important to prove the consistency of the solution. We do
that by replacing (\ref{ft.20}) in (\ref{ft.10}). We assume that $n=-1$. We
find that%

\[
\bar{\rho}_{r0}=\frac{1}{2}\left(  a_{0}\right)  ^{4},~a_{2}=-\frac{7}{8}%
\frac{a_{1}^{2}}{a_{0}}~,~a_{3}=\frac{5}{4}\frac{a_{1}^{3}}{a_{0}}%
~,~a_{4}=-\frac{273}{128}\frac{a_{1}^{4}}{a_{0}}+\frac{a_{0}}{18}%
\Lambda,~\mathit{etc},
\]

where again $\bar{\rho}_{r0}=\frac{1}{2}\left(  a_{0}\right)  ^{4}$ and from
(\ref{ft.09}) we have $16\pi G\rho_{m0}=9a_{0}^{2}a_{1}$.

\paragraph{Case $n>1:$}

When $n>1$ the dominant term in the movable singularity of the field equation
(\ref{ft.10}) follows from the term $\left(  -T\right)  ^{n}$ in the action
and does not correspond to a radiation fluid as occurred in the previous case
with $n<1$. We find that the dominant behaviour is $a\left(  \tau\right)
=a_{0}\tau^{\frac{2}{3}n}$, for $\frac{2}{3}n\notin%
\mathbb{N}
^{\ast}$. Straightforwardly, we calculate the resonances and they are
$s_{1}=-1$, and $s_{2}=0$; that is, the solution is expressed in a right
Painlev\'{e} series where the coefficient $a_{0}$ is the second constant of
integration. This is possible for $n\in%
\mathbb{Q}
$.

Again, when $\frac{2}{3}n=\mu~\in%
\mathbb{N}
^{\ast}$, \ we change variable via $a\left(  \tau\right)  \rightarrow\left(
b\left(  \tau\right)  \right)  ^{-1}$. We find that the field equations pass
the singularity analysis when $\mu$ is an even number,~$\mu=2\zeta$, where the
dominant behaviour is $b\left(  \tau\right)  =b_{0}\tau^{-\frac{3}{2}\zeta}$,
with $\bar{\rho}_{r0}=-2^{-3\zeta}3^{-1+9\zeta}\left(  6\zeta-1\right)
a_{0}^{-4}$ with resonances $s_{1}=-1$ and $s_{2}=\frac{3}{2}\zeta$. Hence the
solution is expressed in a right Painlev\'{e} series in which the step is
$\frac{1}{2}$ for $\zeta$ an odd number and $1$ when $\zeta$ is an even
number. The position of the second constant of integration depends upon the
value of the resonance, $s_{2}$.

\section{Analytical solutions in Bianchi I spacetime}

\label{bianchi1}

The exact solution of the vacuum field equations which follow from the
lagrangian function (\ref{ft.b04}) in GR, \textit{ie} $f\left(  T\right)  =T$,
is the Kasner spacetime where the coefficient functions of the spacetime
(\ref{ft.b02}) are power-law, that is, $\chi\left(  t\right)  =t^{p_{i}}$, and
the $p_{i}=\left(  p_{i},p_{2},p_{3}\right)  $ are solutions of the following
system%
\begin{equation}%
{\displaystyle\sum\limits_{i=1}^{3}}
p_{i}=1~,~%
{\displaystyle\sum\limits_{i=1}^{3}}
p_{i}^{2}=1~. \label{ft.21}%
\end{equation}
These are called Kasner's relations.

However, in modified theories of gravity it is possible for Kasner-like
solutions to exist but Kasner's relations may have to be modified because the
components of the geometric fluids exist. This has been considered first for
the higher-order theories of gravity by Barrow and Clifton in
\cite{barc1,barc2,midd}. Kasner-like solutions have been studied for the
$f\left(  X\right)  =R^{n}$, $f\left(  X\right)  =\left(  R_{\mu\nu}R^{\mu\nu
}\right)  ^{n}~$and $f\left(  X\right)  =\left(  R_{\mu\nu\sigma\lambda}%
R^{\mu\nu\sigma\lambda}\right)  ^{n}$ theories of gravity. Specifically
Kasner's relations (\ref{ft.21}) have been modified such that the right-hand
sides of eq. (\ref{ft.21}) do not equal one, but depend upon the power $n$
defining the lagrangian of the theory, but Kasner's metric or that of
Minkowski spacetime can still be recovered.

Before we study the existence of analytical solutions in the models
$f_{I}\left(  T\right)  $ and $f_{II}\left(  T\right)  $ we consider the
power-law theory $f\left(  T\right)  =\left(  -T\right)  ^{n}$ for which we
study the existence of a Kasner-like solution.

\subsection{Kasner-like solution}

\label{kasnerlike}

Consider $f\left(  T\right)  =\left(  -T\right)  ^{n}$, and assume that%

\[
a\left(  t\right)  =a_{0}t^{p_{1}},~b\left(  t\right)  =b_{0}t^{p_{2}%
}~,~c\left(  t\right)  =c_{0}t^{p_{3}}.
\]

We find that the field equations which follow from the lagrangian
(\ref{ft.b02}) are satisfied either when
\begin{equation}
p_{1}=p_{2}=p_{3}~\text{where }n=\frac{1}{2} \label{ft.22a}%
\end{equation}
~ or when $p_{i}$ satisfies the two conditions%
\begin{equation}%
{\displaystyle\sum\limits_{i=1}^{3}}
p_{i}=2n~-1,~%
{\displaystyle\sum\limits_{i=1}^{3}}
p_{i}^{2}=\left(  2n~-1\right)  ^{2},~\text{for }n>0 \label{ft.22}%
\end{equation}
or
\begin{equation}%
{\displaystyle\sum\limits_{i=1}^{3}}
p_{i}=\rho_{0}~,~%
{\displaystyle\sum\limits_{i=1}^{3}}
p_{i}^{2}=\rho_{0}^{2},~\text{for }n>1 \label{ft.22b}%
\end{equation}

Solution (\ref{ft.22a}) has been derived in \cite{Rod}, but it is that of an
isotropic universe for the theory $f\left(  T\right)  =\sqrt{-T}$, but (as
discussed above) this lagrangian cannot recover the field equations of GR in
an appropriate limit. Furthermore, from (\ref{ft.22}), Kasner's spacetime is
recovered only when $n=1$, while from (\ref{ft.22b}), and for $n>1,$ Kasner's
solution is recovered always for $\rho_{0}=1$. Moreover, there exists
consistency of (\ref{ft.22}) for every value of $n$, while solution
(\ref{ft.22}) has the universe expanding when $n>\frac{1}{2}~$in
(\ref{ft.22}), or $\rho_{0}>0$ in (\ref{ft.22b}), and $t=0$ describes the
position of the spacetime Weyl curvature singularity. Finally, from
(\ref{ft.22}), we observe that for positive values of $n~$(or positive
$\rho_{0}$) one of the resonances always has a different sign from the others,
\textit{ie}, if $p_{2},p_{3}$ are positive, then $p_{1}<0$. \ \ This means
that the chaotic dynamical behaviour on approach to the singularity in the
Mixmaster universe, via an infinite sequence of Kasner eras, can occur as in GR,

We see that\ by rescaling via $\bar{p}_{i}=\frac{1}{2n-1}p_{i}$, or $\bar
{p}_{i}=\frac{1}{\rho_{0}}p_{i},$ conditions (\ref{ft.22}) and (\ref{ft.22b})
simply become the GR Kasner relations,%
\begin{equation}%
{\displaystyle\sum\limits_{i=1}^{3}}
\bar{p}_{i}=1~,~%
{\displaystyle\sum\limits_{i=1}^{3}}
\bar{p}_{i}^{2}=1.~ \label{ft.22c}%
\end{equation}

The existence of solutions (\ref{ft.22}), (\ref{ft.22b}) means that the field
equations in $f\left(  T\right)  $-gravity, for the diagonal frame
(\ref{ft.b01}), admit an anisotropic exact solution. This is contrary to the
claim in a recent review of $f\left(  T\right)  $-gravity \cite{ftRev},~which
is based on results in ref. \cite{Rod}. To see this more clearly, note that
the constraint equation $\tilde{G}_{0}^{0}=0$,~where $\tilde{G}_{~\nu}^{\mu}$
is the modified Einstein tensor, is again equation (\ref{ft.23}) for the case
of vacuum. This admits the general solution, $f\left(  T\right)  =\sqrt{-T}$,
and also the particular solution, $T=0$, with $f\left(  T\right)
|_{T\rightarrow0}=0,$ for the power-law case. It is easy to see that the
solution (\ref{ft.22}) allows (\ref{ft.b03}) to take a zero value.

\subsection{Singularity analysis}

We perform our singularity analyses for the models $f_{I}\left(  T\right)
~$and $f_{II}\left(  T\right)  $. As in the case of the isotropic universe, we
will study the two different cases for which $n<1$ and $n>1$.

\subsubsection{Model $f_{I}\left(  T\right)  $}

\paragraph{Case $n<1:$}

For values of $n<1,$ we find that the dominant term in the field equations is
the linear term in the action; that is, we are in the limit of GR as for the
FLRW universe in the previous section. Hence, the dominant terms are $\left(
a\left(  t\right)  ,b\left(  t\right)  ,c\left(  t\right)  \right)  =\left(
a_{0}t^{p_{1}},b_{0}t^{p_{2}},c_{0}t^{p}\right)  $, where $a_{0},b_{0},c_{0}$
are arbitrary constants and the $p_{i}$ satisfy the Kasner relations
(\ref{ft.21}). However, since the $p_{i}$ satisfy the Kasner relations we have
that $T\left(  t\right)  =0$. Hence the singularity analysis fails.

\paragraph{Case $n>1:$}

In the second case, when $n>1$, \ we find that the dominant terms are $\left(
a\left(  t\right)  ,b\left(  t\right)  ,c\left(  t\right)  \right)  =\left(
a_{0}t^{p_{1}},b_{0}t^{p_{2}},c_{0}t^{p}\right)  $, where again $a_{0}%
,b_{0},c_{0}$ are arbitrary constants and the $p_{i}$ satisfy the modified
Kasner relations (\ref{ft.22}). This solution also gives $T\left(  t\right)
=0$, which means that the singularity analysis fails. However in this case we
observe that Kasner's solution (\ref{ft.21}) solves the field equations

\subsubsection{Model $f_{II}\left(  T\right)  $}

For the second model, namely $f_{II}\left(  T\right)  ,$ the singularity
analysis fails to provide us with a solution because the dominant terms ensure
$T\left(  t\right)  =0$. In contrast to the model $f_{I}\left(  T\right)  $,
we now find $f_{II}\left(  T\right)  \neq0$, which means that the field
equations are not satisfied.

\section{TEGR in nonlinear $f\left(  T\right)  $-gravity}

\label{grsol}

We rewrite the gravitational field equations, (\ref{ft.06}), as follows%
\begin{equation}
e_{i}^{\rho}\mathbf{G}f_{T}+\frac{1}{4}e_{i}^{\rho}\left[  \left(
f-Tf_{T}\right)  \right]  +e_{i}^{\rho}S_{\rho}{}^{\mu\nu}\partial_{\mu}%
({T})f_{TT}=4\pi Ge_{i}^{\rho}\mathcal{T}_{\rho}{}^{\nu}, \label{ft.31}%
\end{equation}
where $\mathbf{G}$ is the Einstein tensor in the teleparallel equivalence
\begin{equation}
e_{i}^{\rho}\mathbf{G=}\left(  e^{-1}\partial_{\mu}(ee_{i}^{\rho}S_{\rho}%
{}^{\mu\nu})-e_{i}^{\lambda}T^{\rho}{}_{\mu\lambda}S_{\rho}{}^{\nu\mu}%
+\frac{1}{4}e_{i}^{\nu}T\right)  . \label{ft.32}%
\end{equation}

Recall that the lagrangian density $T$ is related to the Ricci scalar by
\begin{equation}
T=-R+2e^{-1}\partial_{\nu}\left(  eT_{\rho}^{~\rho\nu}\right)  . \label{ft.33}%
\end{equation}

If $\left(  f-Tf_{T}\right)  =0,~$that is $f\left(  T\right)  =T$ or $f\left(
T\right)  _{|T\rightarrow0}=0$ and $T=0,$ then equation (\ref{ft.31}) becomes%
\begin{equation}
e_{i}^{\rho}\mathbf{G}f_{T}+e_{i}^{\rho}S_{\rho}{}^{\mu\nu}\partial_{\mu}%
({T})f_{TT}=4\pi Ge_{i}^{\rho}\mathcal{T}_{\rho}{}^{\nu}.\label{ft.33a}%
\end{equation}
A vacuum solution of $f(T)$ gravity is therefore also a vacuum solution of GR
if and only if
\begin{equation}
R=2e_{\nu}^{-1}\partial_{\nu}\left(  eT_{\rho}^{~\rho\nu}\right)
=0.\label{ft.34a}%
\end{equation}
However, if we assume a non-zero energy-momentum tensor, $e_{i}^{\rho
}\mathcal{T}_{\rho}{}^{\nu}$, then solution (\ref{ft.33a}) is again one of GR
if $f_{T}\neq0$, $e_{i}^{\rho}S_{\rho}{}^{\mu\nu}\partial_{\mu}({T})f_{TT}%
=0~$and condition $R=2e_{\nu}^{-1}\partial_{\nu}\left(  eT_{\rho}^{~\rho\nu
}\right)  ~$holds. The latter conditions have been derived in \cite{rfa}. \ In
the case of vacuum it is not necessary that $f_{T}~$\ be a non-zero constant.
It can be also zero when GR is recovered as we saw in Section \ref{kasnerlike}
with the case of power-law $f\left(  T\right)  $.

We conclude that vacuum solutions in GR can be recovered in $f\left(
T\right)  $-gravity as in the case of the fourth-order $f\left(  R\right)
$-gravity \cite{bar}. However, it is necessary to select the correct frame in
which $R=2e_{\nu}^{-1}\partial_{\nu}\left(  eT_{\rho}^{~\rho\nu}\right)  $.
Note also that a vacuum solution of GR may correspond only to a special
solution of $f(T)$ gravity and may not be stable in initial data space
\cite{bargen}.

For the Bianchi I model, a power-law solution, $a\left(  t\right)
=a_{0}t^{p_{1}},~b\left(  t\right)  =b_{0}t^{p_{2}}~,~c\left(  t\right)
=c_{0}t^{p_{3}},$ solves the vacuum field equations if $T=0$, $f\left(
T\right)  |_{T\rightarrow0}=0,$ which provides the constraint equation
\begin{equation}
\left(  p_{1}p_{2}+p_{1}p_{3}+p_{2}p_{3}\right)  =0, \label{ft.24}%
\end{equation}
if and only if the lhs of (\ref{ft.33a}) is well defined. In the case of
$f\left(  T\right)  =T+\alpha\left(  -T\right)  ^{n}$, where $f\left(
0\right)  =0$, $f_{T}\left(  0\right)  =1$, we have that $e_{i}^{\rho}S_{\rho
}{}^{\mu\nu}\partial_{\mu}({T})f_{TT}=0$ only when $n>1$ and then GR is recovered.

On the other hand, in $f\left(  T\right)  =\left(  -T\right)  ^{n}$ gravity we
have that $f\left(  0\right)  =0$, $f_{T}\left(  0\right)  =0$ and
$e_{i}^{\rho}S_{\rho}{}^{\mu\nu}\partial_{\mu}({T})f_{TT}=0$ for $n>1$, where
condition (\ref{ft.25}) provides us with (\ref{ft.22b}), where the Kasner
solution is recovered again for $\rho_{0}=1$ without necessarily having
$f_{T}\left(  0\right)  \neq0$. However, for values of $n$ where $n\in\left(
0,1\right)  ,~$the quantities $f_{T}\left(  0\right)  ,~f_{TT}\left(
0\right)  $ are infinite but if the~constant~$\rho_{0}$ has the value
$\rho_{0}=2n-1$, then the rhs part of (\ref{ft.33a}) is well defined.

\subsection{Cosmological constant}

If we include the cosmological constant, then eq. (\ref{ft.31}) becomes%
\begin{equation}
e_{i}^{\rho}\left(  \mathbf{G+\Lambda}\right)  f_{T}+\frac{1}{4}e_{i}^{\rho
}\left[  \left(  f-\left(  T+\Lambda\right)  f_{T}\right)  \right]
+e_{i}^{\rho}S_{\rho}{}^{\mu\nu}\partial_{\mu}({T})f_{TT}=4\pi Ge_{i}^{\rho
}\mathcal{T}_{\rho}{}^{\nu}. \label{ft.25}%
\end{equation}
The above analysis holds and we reduce to the solutions of GR with the
cosmological constant when $f\left(  T\right)  _{|T\rightarrow\Lambda}=0$ and
$T=-\Lambda~$\cite{rfa}. Again, in the vacuum scenario, $f_{,T}\left(
-\Lambda\right)  $ can be zero and GR can be recovered with the proper frame
for the cosmological constant $\Lambda$.

In order to demonstrate this, note that in (\ref{ft.09}), (\ref{ft.10}) and
for the diagonal frame we considered in Section \ref{field}, that $f\left(
T\right)  =\left(  -T-\Lambda\right)  ^{n}$; this means that $f\left(
-\Lambda\right)  =0$, and $f_{,T}\left(  -\Lambda\right)  \rightarrow0$ for
$n>1$ or $f_{,T}\left(  -\Lambda\right)  \rightarrow\infty$ for $n<1$.

From the field equations, (\ref{ft.09}) and (\ref{ft.10}), we find the de
Sitter solutions
\begin{equation}
a\left(  t\right)  =a_{0}\exp\left(  \pm\sqrt{\frac{\Lambda}{6\left(
1-2n\right)  }}t\right)  ~,~n\neq0, \label{ft.28}%
\end{equation}%
\begin{equation}
a\left(  t\right)  =a_{0}\exp\left(  \pm\sqrt{\frac{\Lambda}{6}\left(
1+\Lambda\right)  }t\right)  ~,~n=\frac{1}{2},~\Lambda\neq-1 \label{ft.28a}%
\end{equation}
and%
\begin{equation}
a\left(  t\right)  =a_{0}\exp\left(  \pm\sqrt{\frac{\Lambda}{6}}t\right)
,~n>1. \label{ft.27}%
\end{equation}

The latter is that in which $T=-\Lambda.$ \ This is the solution through which
we recover GR. We observe that (\ref{ft.28}) and (\ref{ft.28a}) provide us
with GR solutions but for a cosmological constant $\tilde{\Lambda}%
=\frac{\Lambda}{1-2n},~\bar{\Lambda}=\Lambda\left(  1+\Lambda\right)  $. This
means that in $f\left(  T\right)  =\left(  -T-\Lambda\right)  ^{n}$ gravity
there exists a solution in which the geometric fluid with components $\rho
_{T},~p_{T}$ has a constant equation of state parameter, $w_{T}=-1$. \ That
follows from the results of\ \cite{rfa} because $f\left(  T\right)  \neq0$ for
$T=\frac{\Lambda}{1-2n}$. \ Then, a new cosmological constant has to be
considered. Recall that for $n=\frac{1}{2}$, the function $f\left(  T\right)
=\left(  -T-\Lambda\right)  ^{\frac{1}{2}}$ is well defined for $\Lambda\neq
0$, in contrast to the situation when $\Lambda=0$.

\ Including a matter source in (\ref{ft.09}), (\ref{ft.10}), like that of a
dust fluid, in order to recover $\Lambda$CDM cosmology we can see that the use
of the condition $T=-\Lambda$ gives the scale factor (\ref{ft.27}), which
means that GR cannot be recovered by that condition -- at least for the frame
that we have considered. We know that $f(T)$-gravity is not invariant under
Lorentz transformations which is one of the main issues with the theory, see
\cite{ftSot,ftTam}. \ Therefore, in order for GR to be recovered,\ the frame
should be that such condition (\ref{ft.34a}) is satisfied.

Consider again the field equations (\ref{ft.09}) and (\ref{ft.10}) without a
matter source~$\rho,~p,$ for a function $f$ such as $T=-\Lambda,~f\left(
-\Lambda\right)  =0$,~with $e_{i}^{\rho}S_{\rho}{}^{\mu\nu}\partial_{\mu}%
({T})f_{TT}=0$. The field equations become%
\begin{equation}
\left(  -T+\Lambda\right)  f_{T}+\left(  f-\left(  T+\Lambda\right)
f_{T}\right)  =16\pi G\rho. \label{ft.30}%
\end{equation}
and%
\begin{equation}
-\left(  4\dot{H}-T+\Lambda\right)  f_{T}-\left(  f-\left(  T+\Lambda\right)
f_{T}\right)  +48H^{2}\dot{H}f_{TT}=16\pi Gp. \label{ft.30a}%
\end{equation}

The de Sitter solution (\ref{ft.27}) solves (\ref{ft.30}), (\ref{ft.30a}) when
$p=-\rho$ and $\rho=\frac{f_{,T}}{8\pi G}\Lambda,~$with $f_{,T}\neq0$, or when
$f_{T}=0$. In the latter case we can say directly that $f\left(  T\right)  $
provides us with the solution of \ the teleparallel equivalence of general
relativity with a cosmological constant in the vacuum, while for $f_{,T}\neq0$
a new\ fluid term has to be introduced in order to eliminate the remaining
terms of $f\left(  T\right)  $ gravity. This is something that is not
necessary when $\Lambda=0$. \ 

Before we close this section we should remark that when $f\left(
-\Lambda\right)  ~$and $f_{T}\left(  -\Lambda\right)  $ are non-zero constants
then the gravitational field equations become those of GR with a cosmological
constant~$\hat{\Lambda}$ which is different to $\Lambda$. Indeed, their
solution will be that of TEGR while we cannot say that GR is always recovered
because of the constraint equation
\begin{equation}
R=2e_{\nu}^{-1}\partial_{\nu}\left(  eT_{\rho}^{~\rho\nu}\right)
+\Lambda\label{ft.30b}%
\end{equation}
.

\section{Conclusions}

\label{conc}

In this paper the method of movable singularities of differential equations
was applied in order to determine analytical solutions of the field equations
in $f\left(  T\right)  $-gravity in a cosmological scenario. \ The models that
we considered are $f_{1}\left(  T\right)  =T+\alpha\left(  -T\right)  ^{n}$
and $f_{2}\left(  T\right)  =T+\alpha\left(  -T\right)  ^{n}-\Lambda~$, where
GR is recovered for $\alpha\rightarrow0$ . For the right hand side of the
field equations, \textit{ie} the energy-momentum tensor, we have considered
two perfect fluids, a dust fluid which corresponds to the cold dark matter and
a blackbody radiation term. We prove that the solution of these models is
given as a right Painlev\'{e} series and the cosmological constant does not
play any significant role in the existence of the movable singularity or on
the resonances. The cosmological constant modifies only the coefficients of
the Painlev\'{e} series$\,$.

We studied two different cases in which the total fluid is (A) dust and (B)
dust plus radiation. For the case (A) we found that the field equations always
pass the singularity test. When $n<1$, the dominant term gives with dust term,
as in GR, while far from the movable singularity, which corresponds to
$a\left(  t_{0}\right)  \rightarrow0,~\dot{a}\left(  t_{0}\right)
\rightarrow\infty$, the term $\alpha\left(  -T\right)  ^{n}-\Lambda$ plays a
dominant role. On the other hand, when $n>1$, the dominant term corresponds to
the $\left(  -T\right)  ^{n}$ term of the action, which provides an effective
perfect fluid with a constant equation of state parameter, $w_{T}=\frac
{n-1}{n}$.

However, the situation is different when we add a radiation fluid. In this
case we showed that, when $n<1$, the dominant behaviour is that of a radiation
fluid in GR. For $n>1$ we have two possible cases. For $n$ such that $\frac
{2}{3}n\notin%
\mathbb{N}
^{\ast}$ the dominant term is that of $\left(  -T\right)  ^{n}$ and, when
$\frac{2}{3}n\in%
\mathbb{N}
^{\ast}$, we found that the field equations pass the singularity test only if
$\frac{2}{3}n$ is a even number. The dominant term is then $a\left(
\tau\right)  =a_{0}\tau^{\frac{1}{2}n}$. Furthermore, for both cases (A) and
(B), the field equations pass the singularity analysis for $n>1~$only if $n$
is a rational number.

We compare our results with the fourth-order gravity defined by the lagrangian
$f_{I}\left(  R\right)  =R+\alpha R^{n}$ that has been studied from the point
of view of the singularity analysis in \cite{palLeach} without a radiation
fluid. There, it was found that the field equations pass the singularity test
when $n$ is a rational number greater than one and the dominant term is that
of the term $R^{n}$ in the lagrangian for $n>1$ with $n\neq\frac{5}{4},2$. Of
course, the two different theories $f\left(  T\right)  =T^{n}$ and $f\left(
R\right)  =R^{m}$ provide power-law solutions. That means that at a level
close to the movable singularity the two different theories,\ $f_{I}\left(
T\right)  ,~f_{I}\left(  R\right)  ~$, provide a similar behaviour for $n,m>1$.

Another issue that deserves comment is that the movable singularity in the
modified Friedmann equation (\ref{ft.10}) for the models studied corresponds
to a spacetime singularity because either (when $a\left(  t_{0}\right)
\rightarrow0$) the Hubble function, the deceleration parameter, or one of
their higher derivatives of the scale factor becomes singular. \ Of course,
that does not mean that the method of movable singularities of differential
equations cannot be applied in cosmological models with no singularities. A
movable singularity~at $t\rightarrow t_{0}$, when it exists, can provide a
solution such as $a\left(  t_{0}\right)  \rightarrow\infty$. That is possible
when the dominant behaviour is negative. This is clear from the analysis we
perform in the Bianchi I spacetime.

When considering the Bianchi I spacetime we found that the vacuum field
equations admit an anisotropic Kasner-like solution which is contrary to the
existing results in the literature \cite{ftRev,Rod}. We did that by studying
the field equations for the power-law model $f\left(  T\right)  =\left(
-T\right)  ^{n}$. The modified Kasner relations depend upon the power, $n$,
and the the sum of the Kasner indices and their squares are $\left(
2n-1\right)  $, and $\left(  2n-1\right)  ^{2}$, or $\rho_{0},~$and, $\rho
_{0}^{2}$, respectively, where for $n=1$ or $\rho_{0}=1$ we are in the limit
of teleparallel equivalence of GR. As far as the two models $f_{I}\left(
T\right)  ,~f_{II}\left(  T\right)  $ are concerned, we found that the
singularity analysis failed to provide us with the analytical solution of the
field equations. However, the dominant terms are also solutions of the field
equations for the $f_{I}\left(  T\right)  $ model, where for $n<1$ the Kasner
solution is recovered, while for $n>1$ the Kasner-like solution follows.
Furthermore, we note that the results are different from that of $f\left(
R\right)  =R^{m}$, gravity, where two families of Kasner-like solutions exist
while the power $m\,$of the theory cannot be arbitrary.

In $f\left(  T\right)  $ gravity for the two spacetimes that we considered we
show that the vacuum field equations are satisfied when the solution
guarantees $T=0$ and $f\left(  T\right)  |_{T\rightarrow0}=0\,.$ In the case
of the FLRW spacetime the solution is that of the four-dimensional Minkowski
spacetime. For the Bianchi I spacetime if we consider a power-law solution
then condition (\ref{ft.24}) should be satisfied and the Kasner metric solves
(\ref{ft.24}). We expect that an $f(T)$ Mixmaster universe to have similar
chaotic behaviour to that displayed in GR on approach to a spacetime singularity.

We also studied when solutions of the teleparallel equivalence of GR can be
recovered in $f\left(  T\right)  $-gravity. We found that when $T=T_{0}$ and
$f\left(  T_{0}\right)  =0$, the field equations do not admit terms which
diverge at infinity. The solution of GR is recovered for the proper frame for
an arbitrary value of $f_{T}\left(  T_{0}\right)  $ for the vacuum case with
or without a cosmological constant, and also when $f_{T}\left(  T_{0}\right)
\neq0$ when a fluid is included in the field equations.

The knowledge that the field equations form an integrable system is important
for the existence of real solutions. Symmetries and singularity analyses are
two independent methods which they provide us with information if the system
is integrable. In a forthcoming work we would like to extend that approach and
in other gravitational theories. 

\begin{acknowledgments}
AP acknowledges financial support \ of FONDECYT grant no. 3160121. JDB
acknowledges support from the STFC. PGL Leach thanks the Instituto de Ciencias
F\'{\i}sicas y Matem\'{a}ticas of the UACh for the hospitality provided while
this work carried out and acknowledges the National Research Foundation of
South Africa and the University of KwaZulu-Natal for financial support. The
views expressed in this paper should not be attributed to either institution.
\end{acknowledgments}

\end{document}